\documentclass[epj]{webofc}
\usepackage[utf8]{inputenc}
\usepackage[varg]{txfonts}   
\usepackage{booktabs}
\usepackage{xcolor}
\definecolor{darkred}{rgb}{0.4,0.0,0.0}
\definecolor{darkgreen}{rgb}{0.0,0.4,0.0}
\definecolor{darkblue}{rgb}{0.0,0.0,0.4}
\usepackage[bookmarks,linktocpage,colorlinks,
    linkcolor = darkred,
    urlcolor  = darkblue,
    citecolor = darkgreen]{hyperref}
\usepackage{ulem}
\wocname{EPJ Web of Conferences}
\woctitle{Lattice2017}
\begin{document}
%
\selectlanguage{english}
\title{%
  \begin{picture}(0,0)(0,0)%
    \put(350,75){\makebox(0,0)[l]{\textnormal{\normalsize RIKEN-QHP-336}}}%
  \end{picture}
  Two-baryon systems from HAL QCD method \\
  and the mirage in the temporal correlation of the direct method
}
\author{%
  \firstname{Takumi} \lastname{Iritani}\inst{1}\thanks{Speaker, \email{takumi.iritani@riken.jp}}  
\lastname{for HAL QCD Collaboration}
}
\institute{%
  Theoretical Research Division, Nishina Center,
  RIKEN, Wako 351-0198, Japan
}
\abstract{%
  Both direct and HAL QCD methods
  are currently used to study the hadron interactions in lattice QCD.
  In the direct method, the eigen-energy of two-particle is measured
  from the temporal correlation.
  Due to the contamination of excited states, however,
  the direct method suffers from the fake eigen-energy problem, 
  which we call  the ``mirage problem,''
  while the HAL QCD method can extract information from all elastic states
  by using the spatial correlation.
  In this work, we further investigate systematic uncertainties of
  the HAL QCD method such as the quark source operator dependence, 
  the convergence of the derivative expansion of the non-local interaction kernel,
  and the single baryon saturation, which are found to be well controlled.
  We  also confirm the consistency between the HAL QCD method
  and the L\"uscher's finite volume formula.
  Based on the HAL QCD potential, we 
  quantitatively confirm that
  the mirage plateau in the direct method 
  is indeed caused by the contamination of excited states.
}
\maketitle
\section{Introduction}\label{intro}

To study the hadron interactions in lattice QCD,
both the direct method \cite{FVM-review} and the HAL QCD method \cite{Ishii:2006ec} are employed.
In the previous studies at heavier quark masses, however,
both dineutron and deuteron are bound in 
the direct method 
\cite{Beane:2011iw,Beane:2012vq,Beane:2013br,Orginos:2015aya,Yamazaki:2011nd, Yamazaki:2012hi,Yamazaki:2015asa,Berkowitz:2015eaa,Wagman:2017tmp},
while they are unbound in the HAL QCD method \cite{Aoki:2012tk, HALQCD:2012aa}.

In the series of papers \cite{Iritani:2015dhu,Iritani:2016jie,Iritani:2017rlk,Aoki:2017byw}, 
we pointed out that the discrepancies come 
from the misidentification of the energy eigenstate (``mirage'' of the true plateau)
in the direct method due to the contamination of excited (scattering) states.
The manifestation of this problem can be exposed by
a ``sanity (consistency) check''~\cite{Iritani:2017rlk,Aoki:2017byw}
using the L\"uscher's formula \cite{Luscher:1991}
and source/sink operator dependences of the plateaux \cite{Iritani:2015dhu,Iritani:2016jie,Iritani:2017rlk}.
These symptoms in the previous studies by the direct method 
\cite{Beane:2011iw,Beane:2012vq,Beane:2013br,Orginos:2015aya,Yamazaki:2011nd,
Yamazaki:2012hi,Yamazaki:2015asa,Berkowitz:2015eaa,Wagman:2017tmp}
cast serious doubt on their conclusions.

In this work, we investigate the reliability of the HAL QCD method,
and show that systematic uncertainties are under control.
We also reveal the origin of the fake plateau in the temporal correlator
quantitatively, and 
demonstrate that correct plateaux emerge for both the ground and the 1st
excited states  if temporal correlation functions are projected to eigenstates
of the HAL QCD potential.

\section{Time-dependent HAL QCD method}
\subsection{Formalism}
In the time-dependent HAL QCD method \cite{HALQCD:2012aa},
one measures the Nambu-Bethe-Salpeter  correlation function given by
\begin{equation}
  R(\vec{r},t)  \equiv \frac{\langle 0 | T \left\{
    B(\vec{x}+\vec{r},t)B(\vec{x},t)\right\}
  \overline{\mathcal{J}}(0)|0 \rangle}{
    \left\{G_B(t)\right\}^2}
    = \sum_n A_n \psi_{W_n}(\vec{r})e^{-(W_n - 2m_B)t} + \mathcal{O}\left(e^{-(W_\mathrm{th}-2m_B)t}\right)
  \label{eq:R-corr}
\end{equation}
with a source operator $\mathcal{J}$,
$n$-th energy eigenvalue $W_n$, the inelastic threshold $W_\mathrm{th}$, single baryon correlator $G_B(t)$ and the baryon mass $m_B$.
With elastic saturation, $R(\vec{r},t)$ satisfies 
\begin{equation}
  \left[ \frac{1}{4m_B}\frac{\partial^2}{\partial t^2}
  - \frac{\partial}{\partial t} - H_0\right]R(\vec{r},t)
  = \int d\vec{r'} U(\vec{r},\vec{r'}) R(\vec{r'},t)
  \label{}
\end{equation}
where $U(\vec{r},\vec{r'})$ is the non-local interaction kernel.
Using the velocity expansion in the spin-singlet S-wave channel,
$U(\vec{r},\vec{r'}) \simeq   V_\mathrm{eff}(r)\delta(\vec{r}-\vec{r'})$,
the effective leading order (central) potential is defined by
\begin{equation}
  V_\mathrm{eff}(r) = \frac{1}{4m_B} \frac{(\partial/\partial t)^2R(\vec{r},t)}{R(\vec{r},t)}
  - \frac{(\partial/\partial t)R(\vec{r},t)}{R(\vec{r},t)}
  - \frac{H_0 R(\vec{r},t)}{R(\vec{r},t)}.
  \label{}
\end{equation}
Considering the higher order term as $U(\vec{r},\vec{r'})\simeq 
\{ V_{LO}(r) + V_{NLO}(r)\nabla^2 \} \delta(\vec{r}-\vec{r'})$,
it leads to 
\begin{equation}
  \frac{1}{4m_B} \frac{\partial^2 R(\vec{r},t)}{\partial t^2}
  - \frac{\partial R(\vec{r},t)}{\partial t}
  - H_0 R(\vec{r},t) = V_{LO}(r)R(\vec{r},t) + V_{NLO}(r)\nabla^2 R(\vec{r},t),
\end{equation}
where the leading order ($V_{LO}(r)$) and next leading order ($V_{NLO}(r)$) potentials
are obtained by solving linear equations with several $R(\vec{r},t)$.

\subsection{Source dependence of HAL QCD method and the next leading order potential}

First, we discuss the quark source dependence of the HAL QCD method.
We use 2+1 flavor QCD configurations in Ref.~\cite{Yamazaki:2012hi},
which are the Iwasaki gauge action and $\mathcal{O}(a)$-improved Wilson quark action
at $a = 0.08995(40)$ fm, where $m_\pi = 0.51$ GeV, $m_N = 1.32$ GeV, and $m_\Xi = 1.46$ GeV.
We employ both the wall source $q^\mathrm{wall}(t) = \sum_{\vec{y}}q(\vec{y},t)$
and the smeared source $q^\mathrm{smear}(\vec{x},t) = \sum_{\vec{y}}f(|\vec{x}-\vec{y}|)q(\vec{y},t)$
with $f(r) \equiv Ae^{-Br}$,\ $1$,\ $0$ for $0 < r < (L-1)/2$,\ $r = 0$,\ $(L-1)/2 \leq r$, respectively,
whose parameters $A, B$ are the same as those in Ref.~\cite{Yamazaki:2012hi}.
The number of the configurations and simulation parameters are summarized in Table~\ref{tab:conf}.

\begin{table}
  \centering
  \caption{The gauge configurations and parameters.}
  \label{tab:conf}
  \begin{tabular}{ccccc}
    \hline
    volume & \# of conf. & \# of smeared source & $(A,B)$ & \# of wall sources \\
    \hline
    $40^3 \times 48$ & 207 & $512$ & (0.8, 0.22) & $48$ \\
    $48^3 \times 48$ & 200 & $4\times 384$ & (0.8, 0.23) & $4\times 48$ \\
    $64^3 \times 64$ & 327 & $1\times 256$ & (0.8, 0.23) & $4\times 64$ \\
    \hline
  \end{tabular}
\end{table}

\begin{figure}
  \centering
  \includegraphics[width=0.47\textwidth,clip]{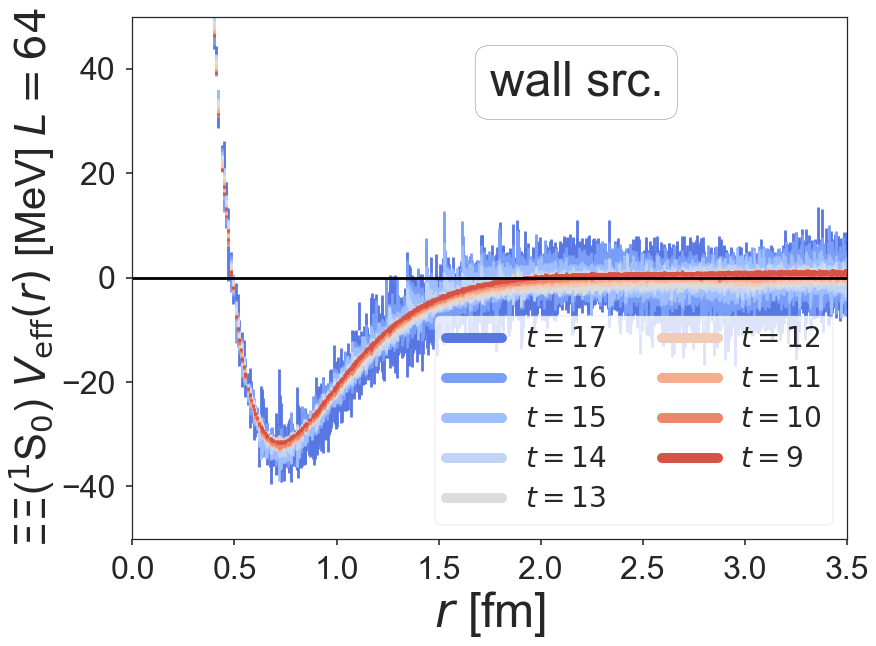}
  \includegraphics[width=0.47\textwidth,clip]{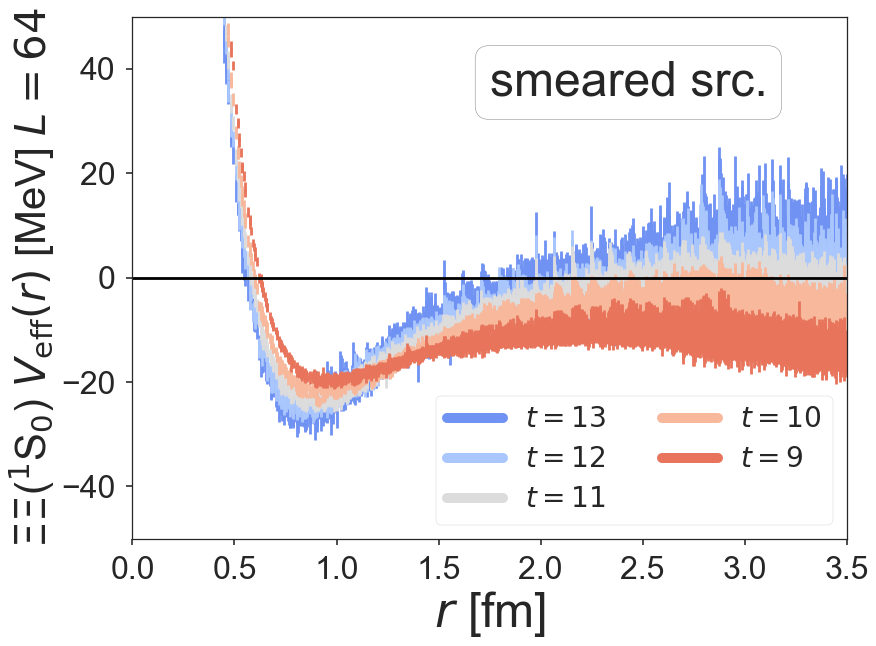}
  \includegraphics[width=0.47\textwidth,clip]{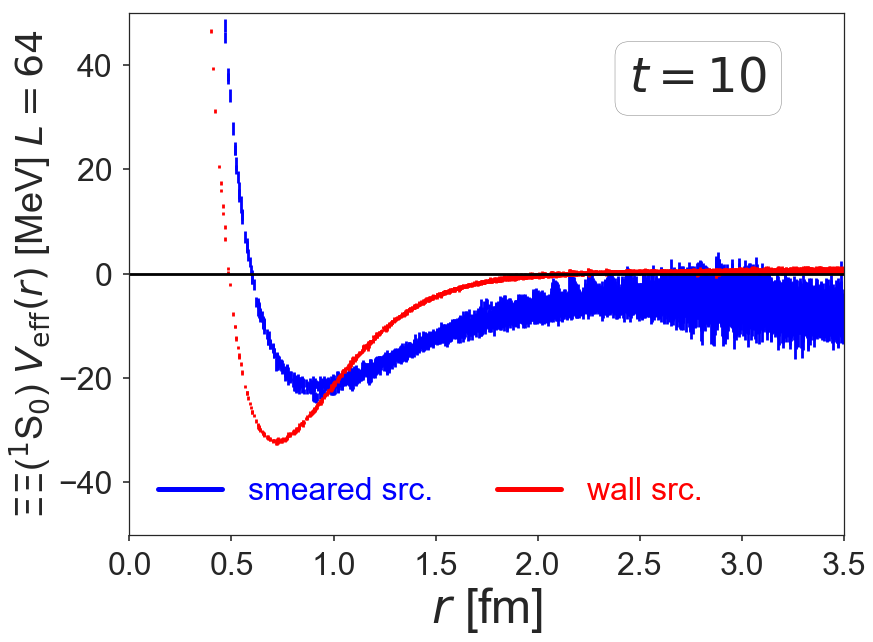}
  \includegraphics[width=0.47\textwidth,clip]{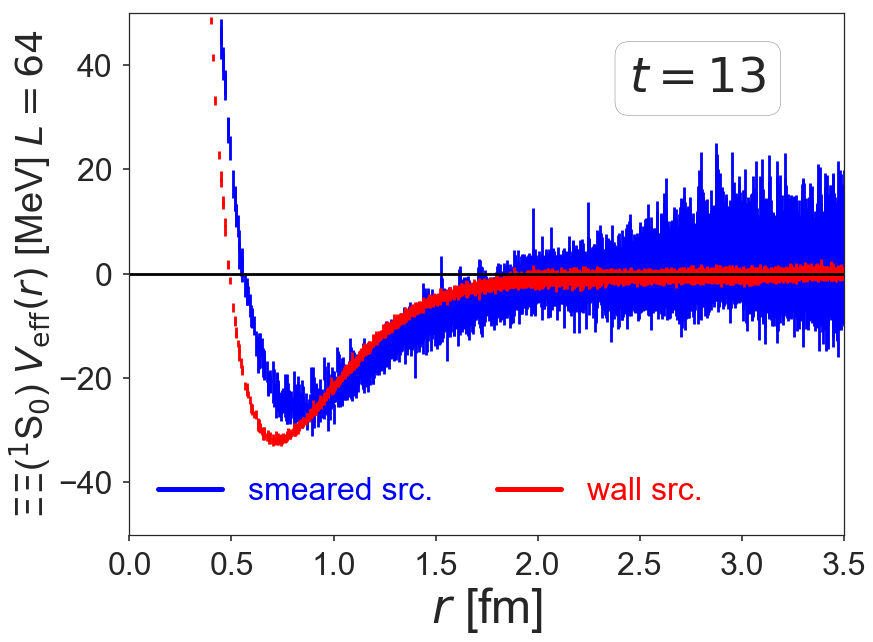}
  \caption{
     The effective leading order potential $V_\mathrm{eff}(r)$ from the wall source and the smeared source at various $t$.
    \label{fig:veff}
  }
\end{figure}

In this work, we focus on $\Xi\Xi$($^1$S$_0$) channel,
which has smaller statistical errors and belongs to the same representation as $NN$($^1$S$_0)$
in the flavor SU(3)  limit.
The upper panels in Fig.~\ref{fig:veff}
show the effective leading order potential  $V_\mathrm{eff}(r)$
from the wall and smeared sources at $L = 64$, respectively.
For the wall source, the potentials are almost unchanged from $t = 9$ to $t=17$, 
while the results from the smeared source  show significant $t$ dependence.
The lower panels in Fig.~\ref{fig:veff} are comparisons between the two  at $t=10$ and 13.
The results imply that
$V_\mathrm{eff}^\mathrm{smear}(r)$ tends to approach to $V_\mathrm{eff}^\mathrm{wall}(r)$
as $t$ increases, while there remain small discrepancies even at $t = 13$.

\begin{figure}
  \centering
  \includegraphics[width=0.47\textwidth,clip]{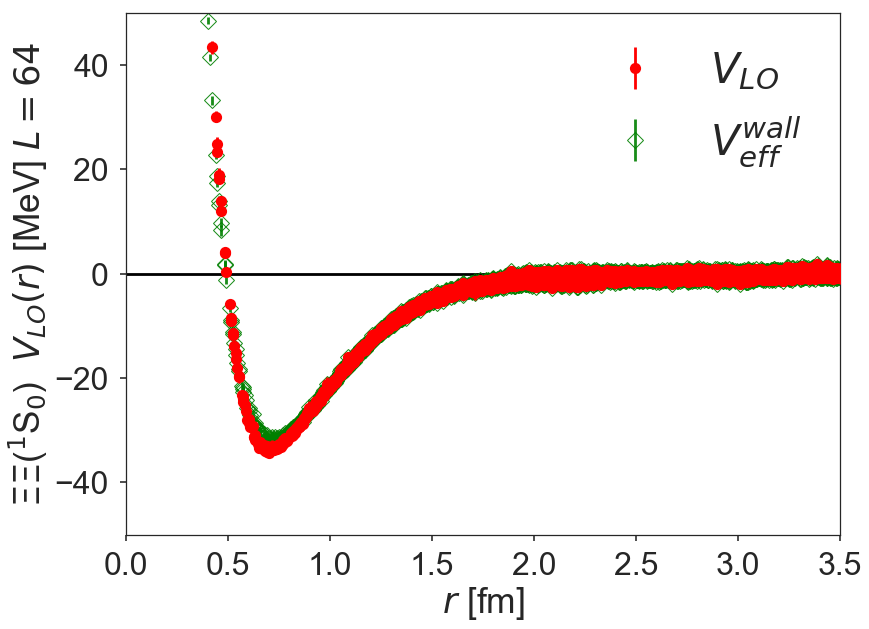}
  \includegraphics[width=0.47\textwidth,clip]{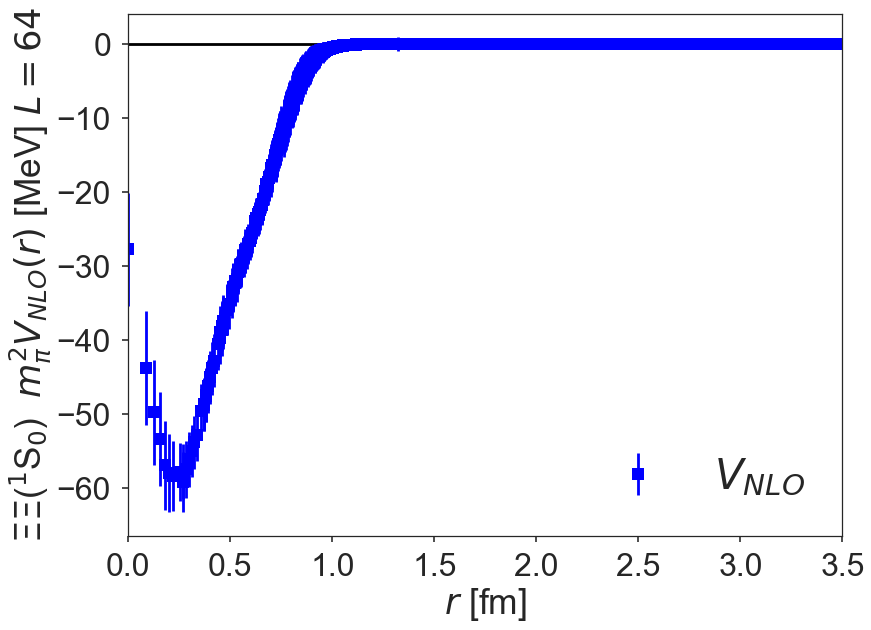}
  \caption{
    \label{fig:vnlo}
    (Left) The effective leading order potential of the wall source,
    and the leading order potential.
    (Right) The next leading order potential.
    }
\end{figure}

The small difference between $V_\mathrm{eff}^\mathrm{wall}(r)$ and $V_\mathrm{eff}^\mathrm{smear}(r)$
indicates the existence of
the next leading order correction in the derivative expansion of the non-local kernel $U(\vec{r},\vec{r'})$.
Fig.~\ref{fig:vnlo} shows the (next) leading order potential $V_\mathrm{LO}(r)$ ($V_\mathrm{NLO}(r)$),
which are obtained by using  $R^\mathrm{wall}(\vec{r},t)$ and $R^\mathrm{smear}(\vec{r},t)$.
The effective leading potential  from the wall source
is almost identical with the leading order potential as shown in Fig.~\ref{fig:vnlo}~(Left),
while in the smeared source,
the next leading order correction 
to the potential, $[V_\mathrm{NLO}(r)\nabla^2 R(\vec{r},t)]/R(\vec{r},t)$, cannot be neglected.

\begin{figure}
  \centering
  \includegraphics[width=0.47\textwidth,clip]{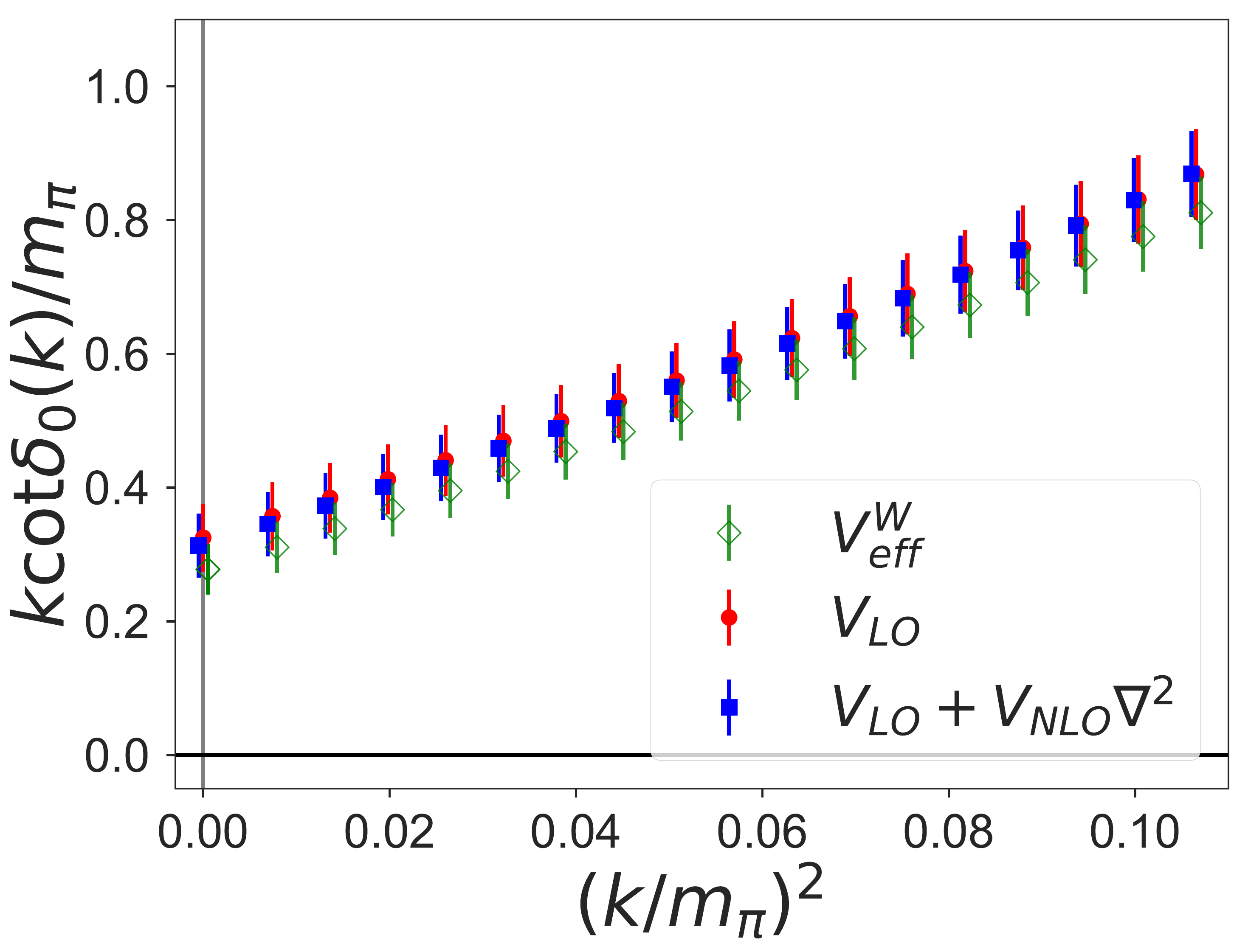}
  \includegraphics[width=0.47\textwidth,clip]{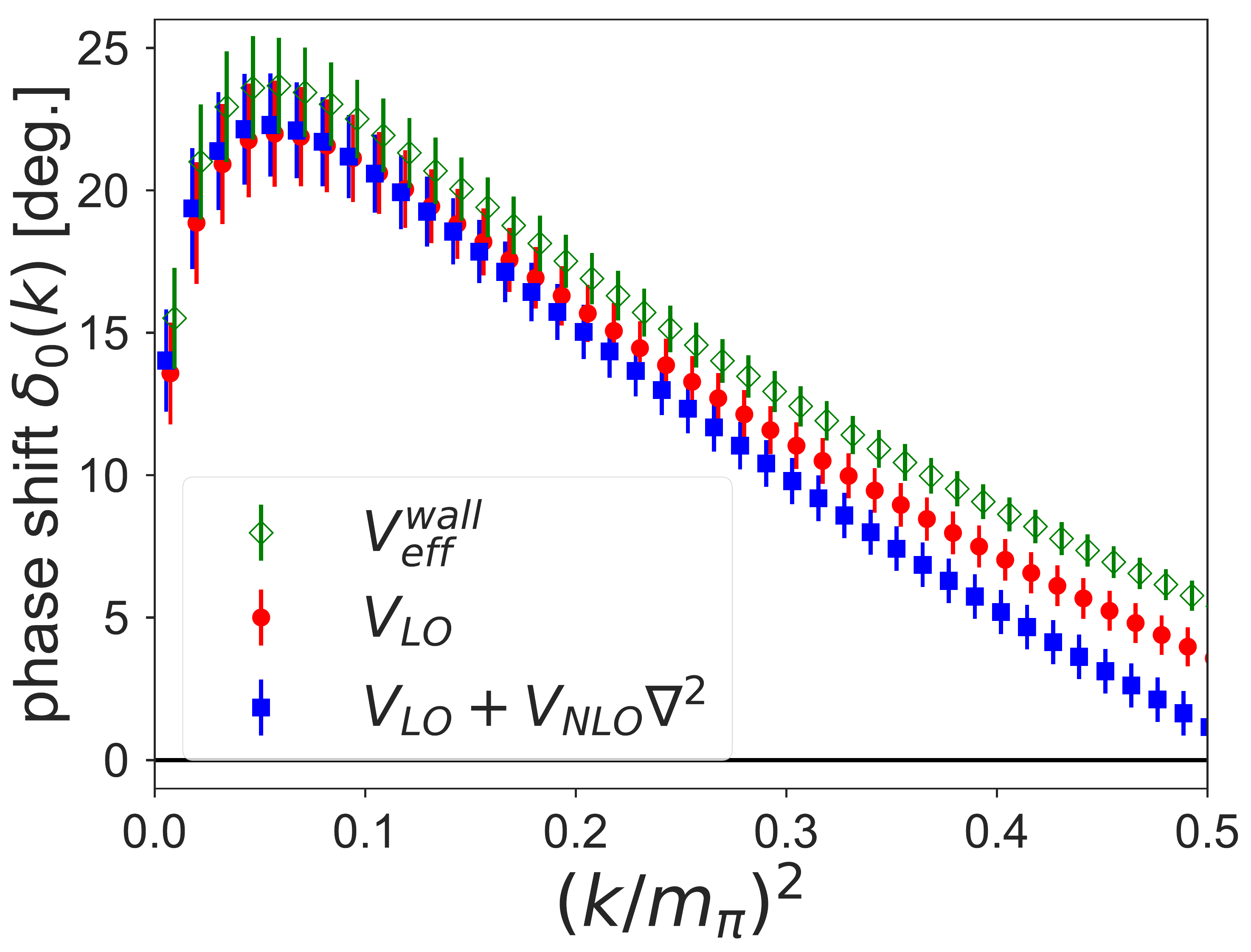}
  \caption{
    \label{fig:scat}
    The scattering phase shifts 
    using $V_\mathrm{eff}^\mathrm{wall}$,
    $V_\mathrm{LO}$ and $V_\mathrm{LO}$ + $V_\mathrm{NLO}\nabla^2$.
    (Left) $k\cot\delta_0(k)$
    (Right) $\delta_0(k)$.
    }
\end{figure}

Fig.~\ref{fig:scat} shows the scattering phase shifts
using $V_\mathrm{eff}^\mathrm{wall}(r)$, $V_\mathrm{LO}(r)$,
and $V_\mathrm{LO}(r) + V_\mathrm{NLO}(r) \nabla^2$.
These phase shifts suggest that $\Xi\Xi(^1$S$_0)$ is an attractive but an unbound channel at $m_\pi = 0.51$ GeV.
As shown in Fig.~\ref{fig:scat}~(Left),
at lower energies, these potentials give the consistent results within statistical error.
The NLO correction  appears only at higher energies (see Fig.~\ref{fig:scat}~(Right)).
These results show
  that (i) the derivative expansion of the non-local kernel has good convergence
  and the corresponding systematic uncertainty can be controlled
  (ii) the effective leading order potential from the wall source
  is reliable at low energies in this system.

\begin{figure}
  \centering
  \includegraphics[width=0.47\textwidth,clip]{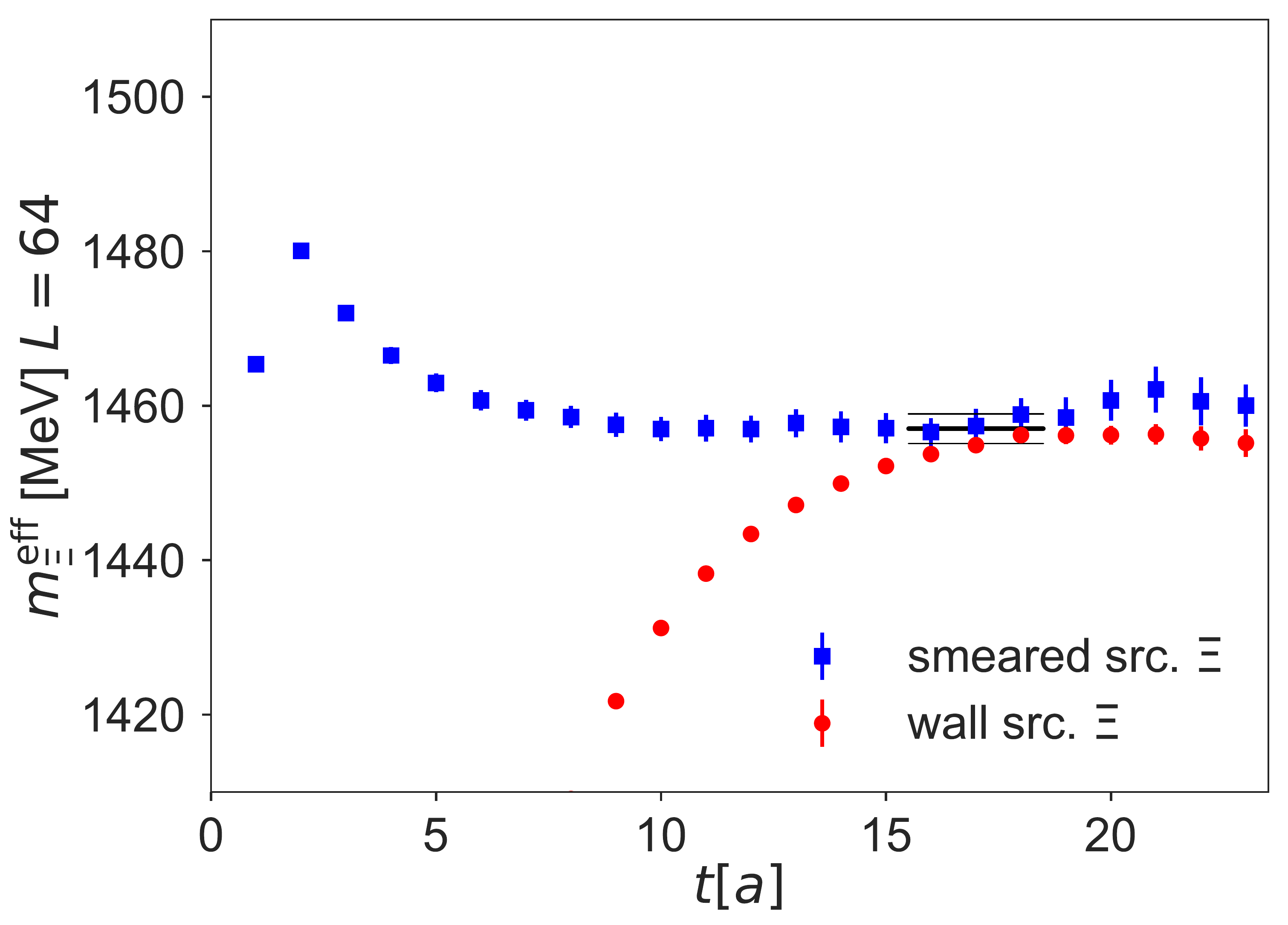}
  \includegraphics[width=0.47\textwidth,clip]{figs/xixi_L64_wall.png}
  \caption{
    \label{fig:eff}
    (Left) The effective mass of the single baryon ($\Xi$).
    (Right) The effective leading order potential of the wall source from $t = 9$ to $t = 17$.
    }
\end{figure}

The smeared source is tuned to have a large overlap with a single baryon ground state,
while the saturation of the single baryon state for the wall source is known to be relatively slower
than that of the smeared source~\cite{Iritani:2016jie}.
Recently, some concerns are expressed to the wall source
for the study of the two-baryon systems \cite{Yamazaki:2017euu,Savage:2016egr}\footnote{
  We note that the validity of the wall source is independent of
    the correctness of the measurement from the smeared source.
    In addition, even within the smeared source, a good ground state saturation
    in a single baryon correlator does not guarantee at all a good ground state saturation
    in a two-baryon correlator.
}.
Fig.~\ref{fig:eff}~(Left) shows the effective masses of the single baryon
from the smeared and the wall sources.
Although the ground state saturation of the wall source 
is slower than that of the smeared source,
the results from both sources converge 
around $t \gtrsim 16$.
(Even at a much earlier time, $t=10$, the difference of the effective mass between the two is as small as 2\%.)
As shown in Fig.~\ref{fig:eff}~(Right) (the same figure as in Fig.~\ref{fig:veff}),
we confirm that 
the wall source potentials at different $t$ are consistent with each other
including the time slices at $t \gtrsim 16$,
and thus the systematic errors from the single baryon saturation
are well under control.
This also indicates that the contaminations from the 
single baryon excited states are almost canceled in the potential at early time slices.

\subsection{Consistency between the L\"uscher's finite volume method and the HAL QCD method}

\begin{figure}
  \centering
  \includegraphics[width=0.47\textwidth,clip]{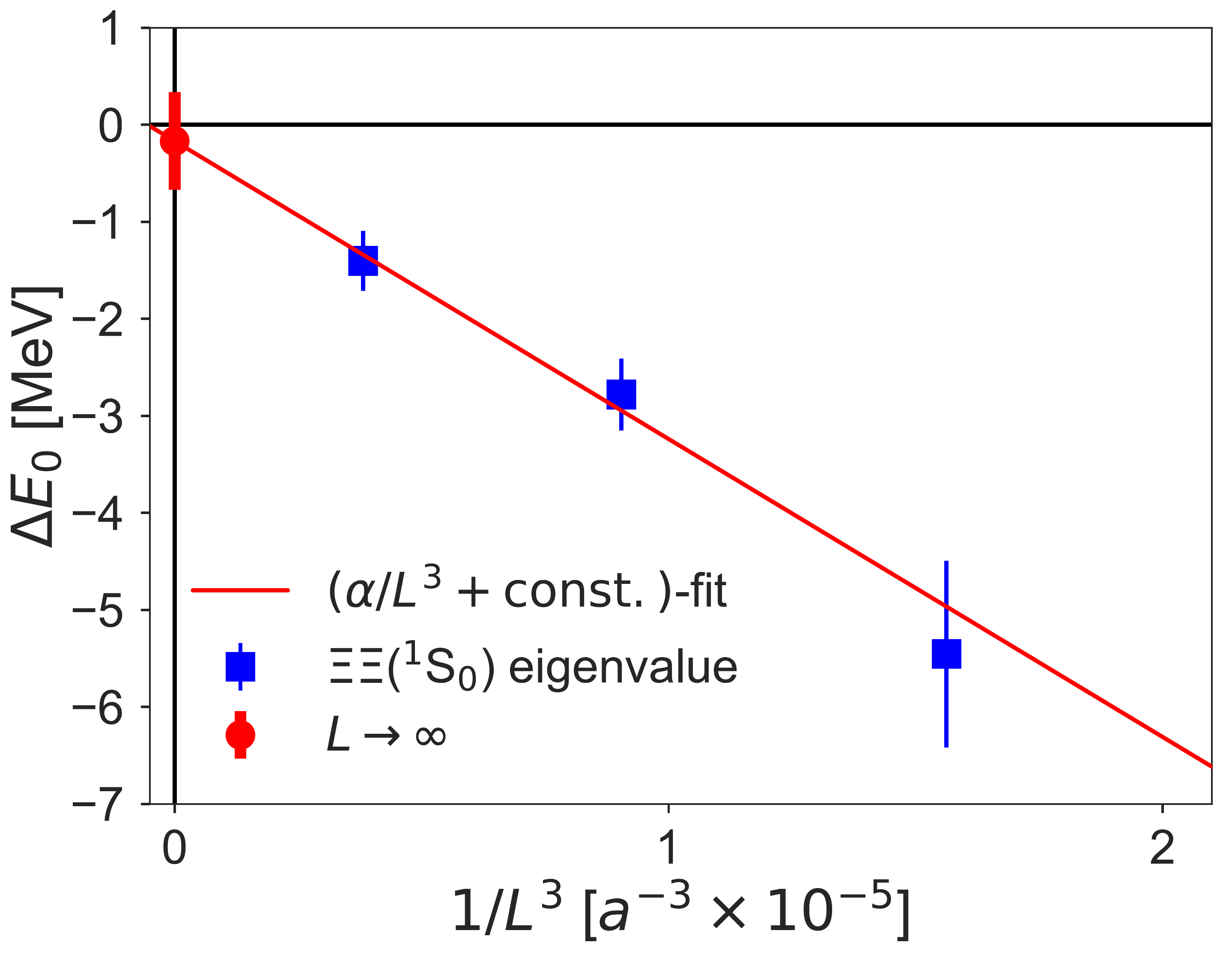}
  \includegraphics[width=0.47\textwidth,clip]{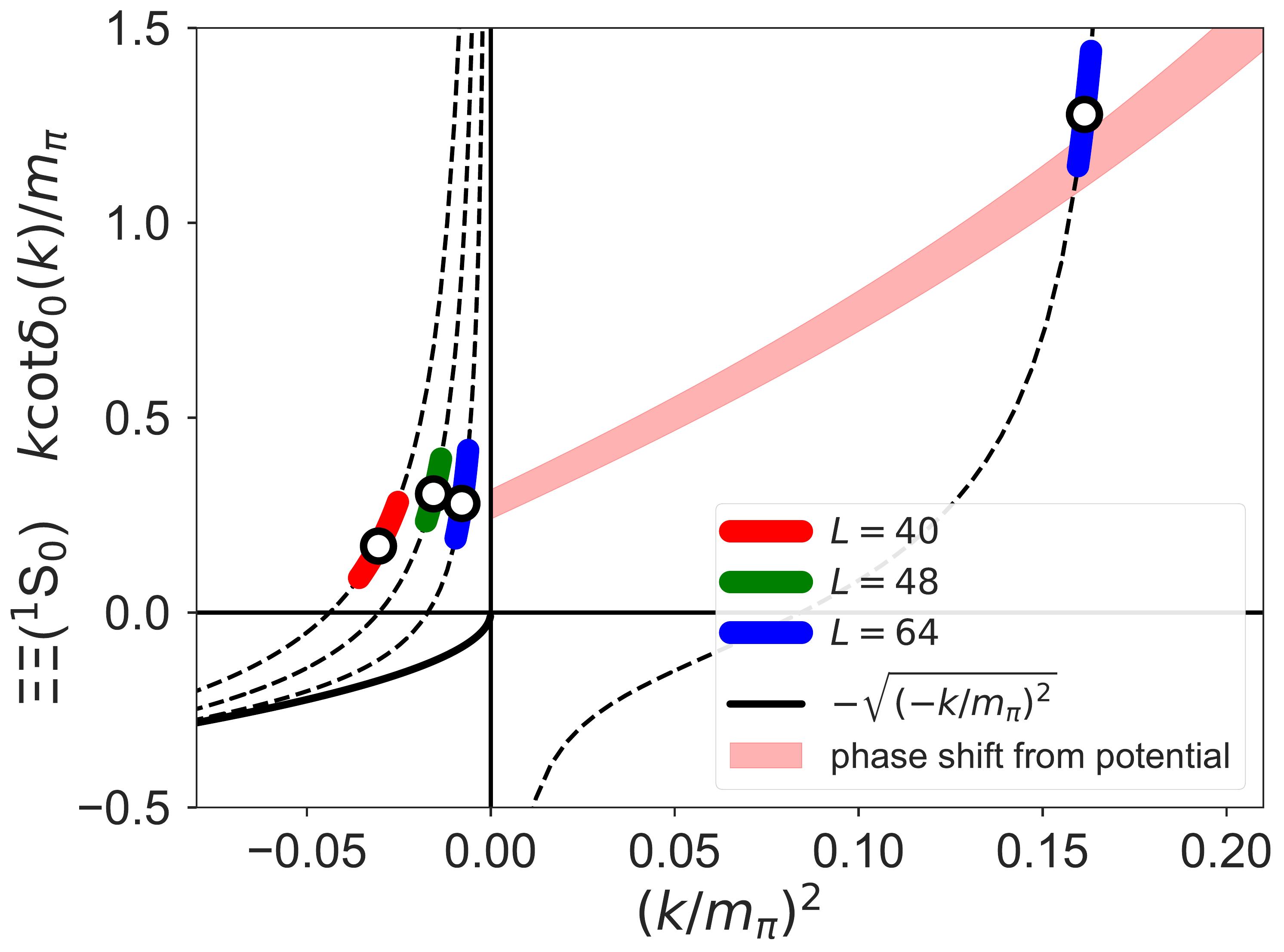}
  \caption{ \label{fig:phase}
    (Left) Volume dependence of the ground state eigen-energy.  
    (Right) The scattering phase shift using the L\"uscher's formula,
    and the HAL QCD potential.
  }
\end{figure}

Next, we discuss the consistency between the HAL QCD potential
and the L\"uscher's finite volume formula \cite{Luscher:1991},
which extracts the scattering phase shift from the energy shift in the finite box.
Fig.~\ref{fig:phase}~(Left)
shows the volume dependence of 
the lowest eigenvalue of the finite volume Hamiltonian $H = H_0 + V_\mathrm{eff}^\mathrm{wall}(r)$.
These spectra are proportional to $1/L^3$ and converge to zero within error
in the infinite volume limit.
This volume dependence 
  of the lowest energy strongly supports an absence of the bound state,
which is consistent with the phase shift analysis by the HAL QCD method in Fig.~\ref{fig:scat}.

We next calculate the scattering phase shift using the L\"uscher's finite volume formula
\begin{equation}
  k\cot\delta_0(k) = \frac{1}{\pi L} \sum_{\vec{n}\in \mathbf{Z}^3}
  \frac{1}{|\vec{n}|^2 - (kL/2\pi)^2},
  \label{}
\end{equation}
where $k$ is given by $\Delta E_L = 2\sqrt{k^2 + m_B^2} - 2m_B$.
Fig.~\ref{fig:phase}~(Right) 
shows $k\cot\delta_0(k)$ as a function of $k^2$ 
using the ground state energy on three volumes and the 1st excited state energy on $L = 64$,
which are compared with $k\cot\delta_0(k)$ in the infinite volume calculated from the HAL QCD potential (pink band).
We here confirm not only a consistency between the two methods 
but also a smooth behavior of the finite volume energy:
$k\cot\delta_0(k)$ for $k^2 > 0$ from the finite volume energy agrees with the pink band
from the potential, and
$k\cot\delta_0(k)$ for $k^2 < 0$ by the L\"uscher's formula 
smoothly converges to the positive intersect at $k^2=0$, consistent with the pink band.

\section{Diagnosis of the direct method form the HAL QCD potential}

\begin{figure}
  \centering
  \includegraphics[width=0.49\textwidth,clip]{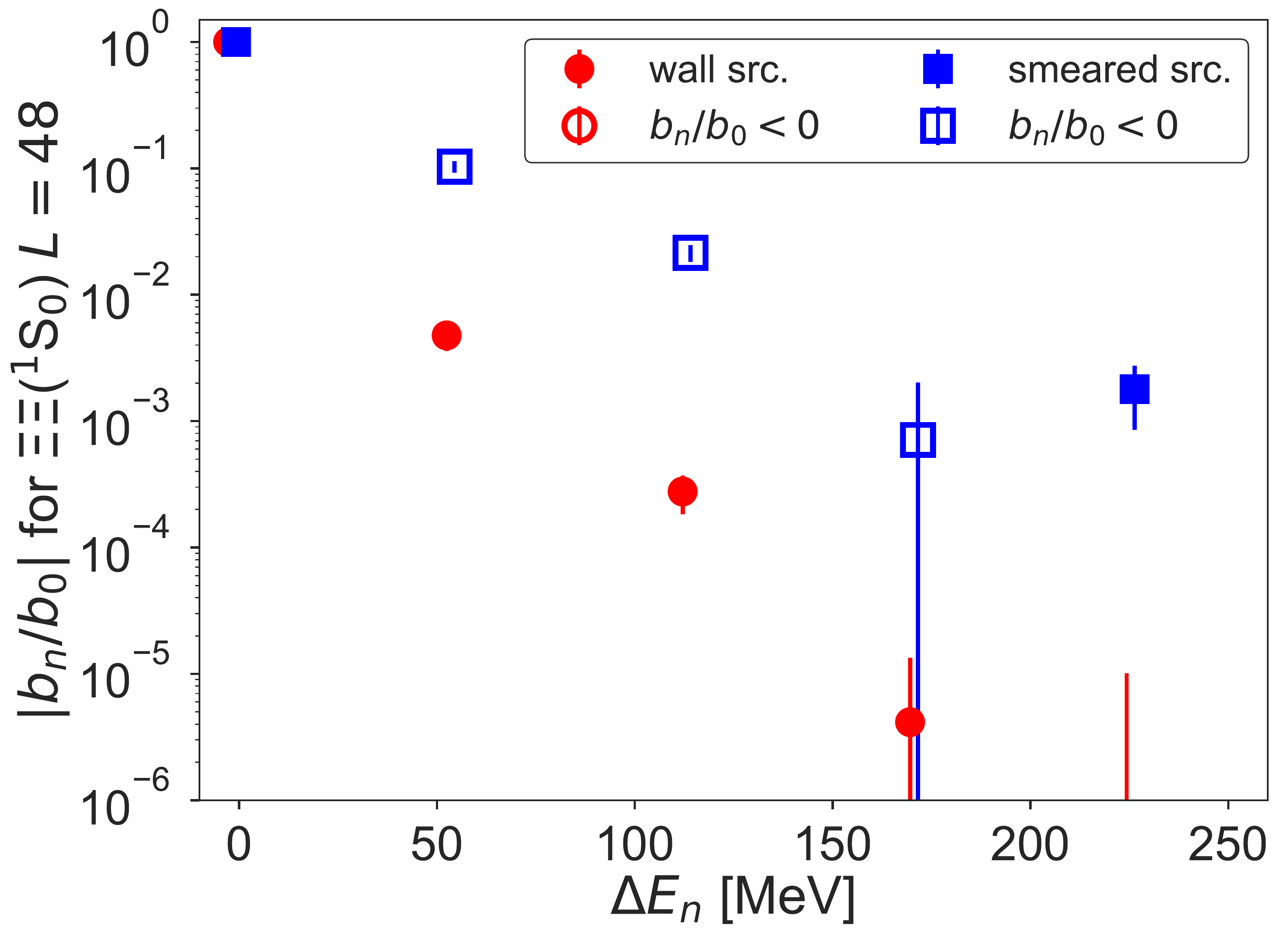}
  \includegraphics[width=0.49\textwidth,clip]{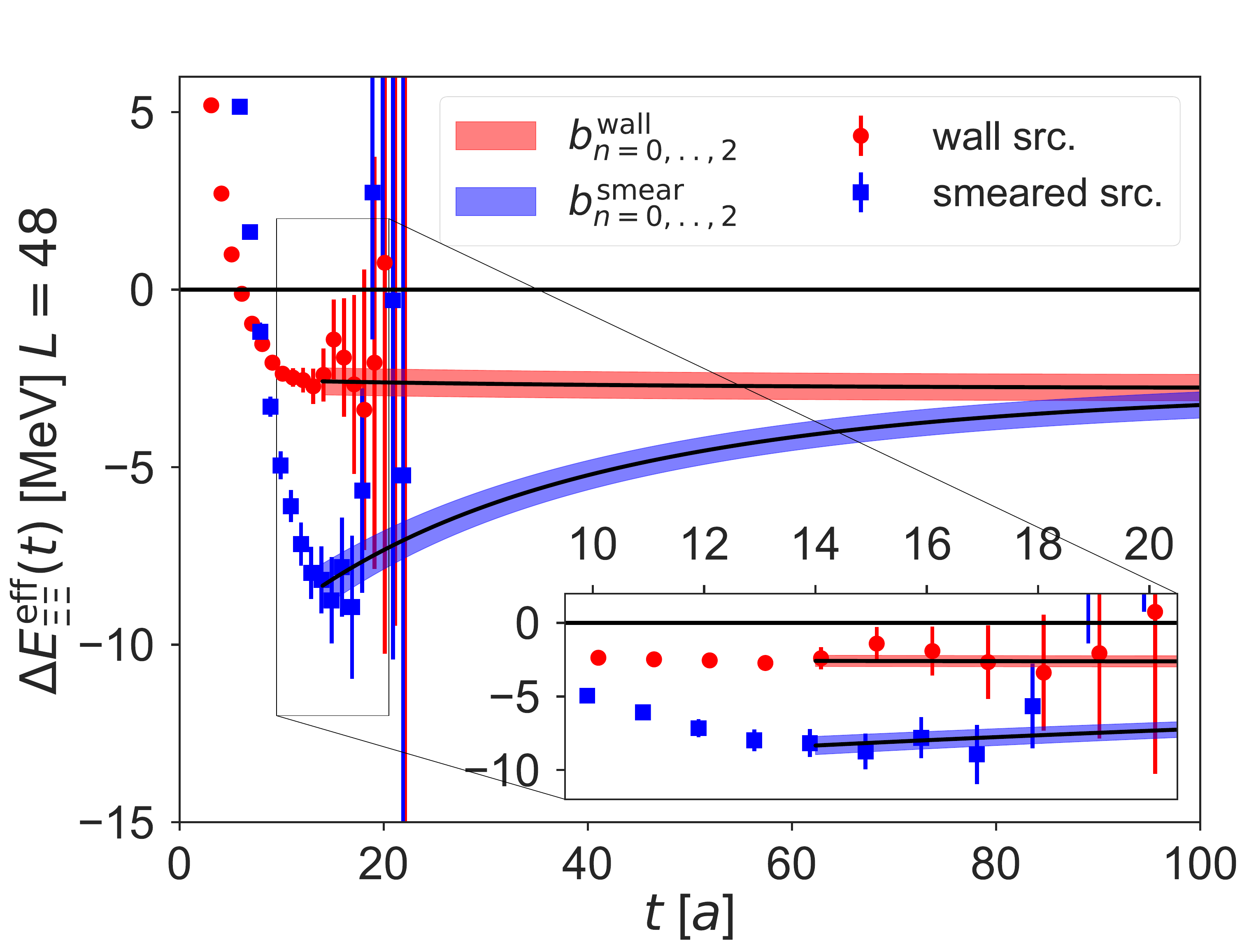}
  \caption{
    \label{fig:bn_b0}
    (Left) The contamination of the excited state $|b_n/b_0|$.
    Open symbols denote negative values ($b_n/b_0 < 0$).
    (Right) The reconstructed effective energy shifts using theree low-lying eigenstates.
  }
\end{figure}

Finally, we reveal the origin of the fake plateau in the direct method.
Using the low-lying eigenfunctions $\Psi_n(\vec{r})$ and eigenvalues $\Delta E_n$,
which are obtained by solving $H = H_0 +  V_\mathrm{eff}^\mathrm{wall}(r)$ in the finite box,
the $R$-correlator in Eq.~(\ref{eq:R-corr}) can be decomposed as
\begin{equation}
  \sum_{\vec{r}}R^\mathrm{wall/smear}(\vec{r},t)
  \simeq \sum_{\vec{r}}\sum_n a_n^\mathrm{wall/smear}
  \Psi_n(\vec{r}) \exp(-\Delta E_n t)
  = \sum_n b_n^\mathrm{wall/smear} \exp(-\Delta E_n t)
  \label{}
\end{equation}
where $a_n^\mathrm{wall/smear}$ is determined by the orthogonality of $\Psi_n(\vec{r})$.
Fig.~\ref{fig:bn_b0}~(Left)
shows the magnitude of the ratio $|b_n/b_0|$ for both wall and smeared sources
as a function of $\Delta E_n$, where the filled (open) symbol
represents a positive (negative) value.
This quantity represents the magnitude of the contamination of
the excited states in $R$-correlator.
For example, the contamination of the 1st excited state is smaller than 1\% in the wall source,
while it is as large as 10\% (with a negative sign) in the smeared source.

In Fig.~\ref{fig:bn_b0}~(Right),
we show the reconstructed effective energy shift using three
low-lying modes for both wall and smeared sources,
which is given by 
\begin{equation}
  \overline{\Delta E}_\mathrm{eff}^\mathrm{wall/smear}(t) = \log 
  \frac{ \sum_{n=0}^{2}b_n^\mathrm{wall/smear} \exp(-\Delta E_n t)}
  { \sum_{n=0}^{2}b_n^\mathrm{wall/smear} \exp(-\Delta E_n (t+1))}.
  \label{}
\end{equation}
It well reproduces 
the (fake) plateaux-like behavior in the direct method around $t = 15$.
We can estimate that about $t \sim $ $100a \sim 10$ fm is required for 
the smeared source to reach the correct ground state.  

\begin{figure}
  \centering
\includegraphics[width=0.48\textwidth,clip]{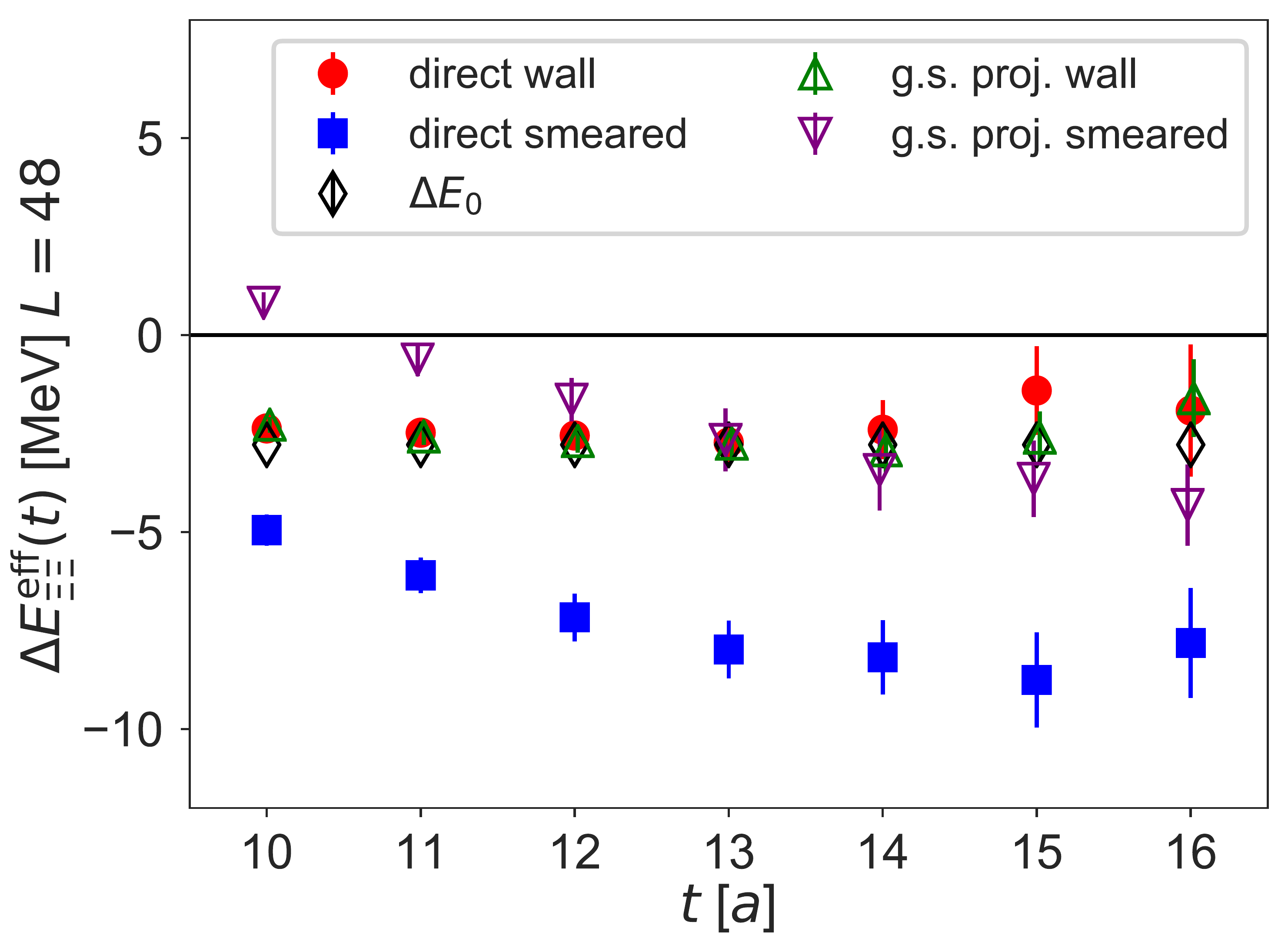}
\includegraphics[width=0.47\textwidth,clip]{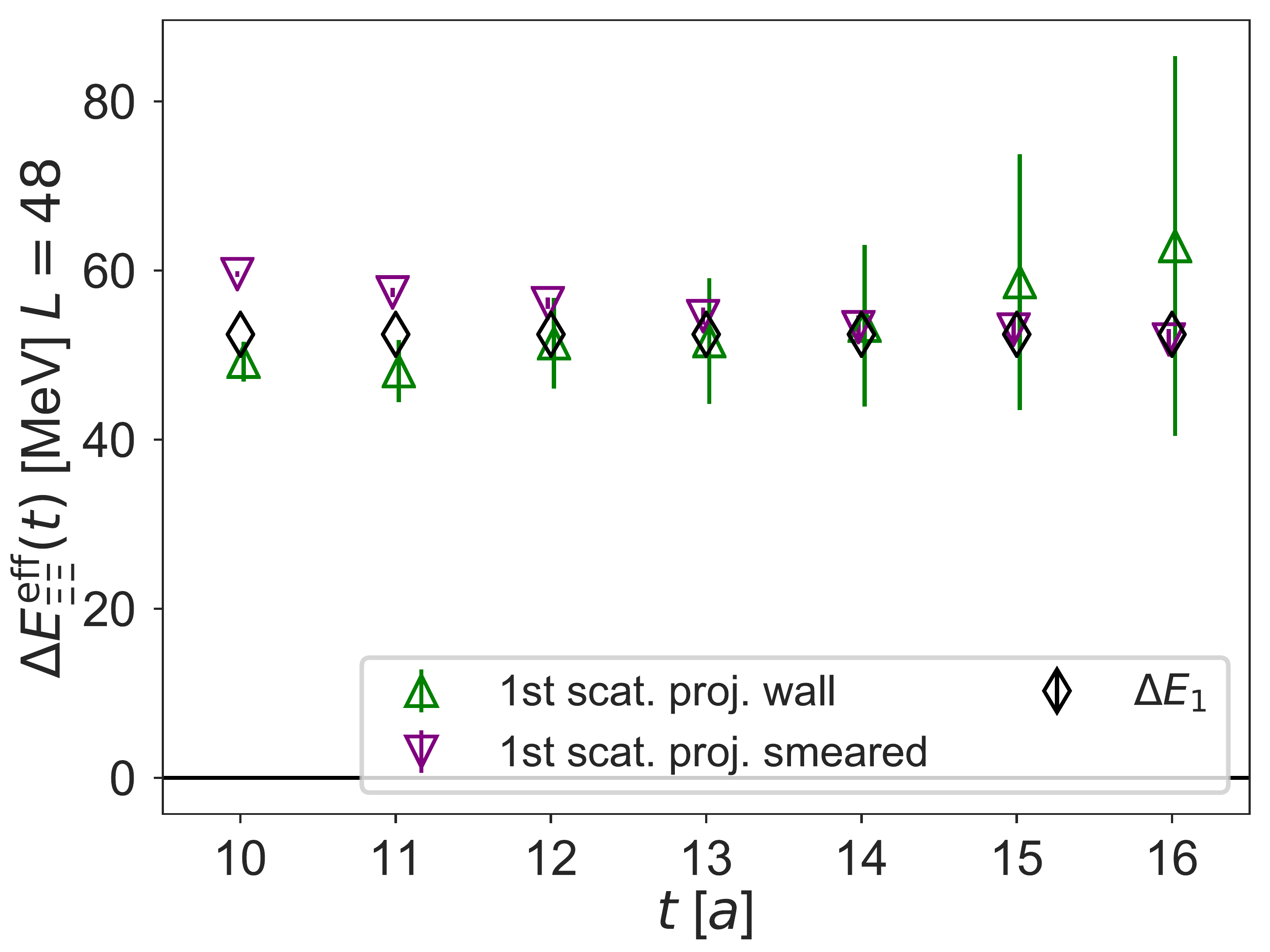}
  \caption{
    \label{fig:eproj}
    The effective energy shift using the ground state (left) and 1st excited state (right)
    projected correlators.
  }
\end{figure}

Finally, we demonstrate reliability of eigenstates of the HAL QCD potential,
using
the projected effective energy shift defined by
\begin{equation}
  \Delta E_\mathrm{eff}^{(n)} = \log \frac{R^{(n)}(t)}{R^{(n)}(t+1)},
  \label{}
\end{equation}
where $R^{(n)}(t) \equiv \sum_{\vec{r}} \Psi_n(\vec{r}) R(\vec{r},t)$ 
with the eigenfunction $\Psi_n(\vec{r})$.
Fig.~\ref{fig:eproj} shows 
projected effective energy shifts for the ground and 1st excited states,
which give source independent plateaux consistent with eigenenergies $\Delta E_{0,1}$ within statistical errors.
This demonstration establishes the correctness of the HAL QCD potential in this case,
since its (finite volume) eingenenergies are faithful to the finite volume energies. 
Moreover, once correct eigenstates are obtained from the potential, we can
construct correlation functions projected to these eigenstates, whose plateaux
agree with correct eigenenergies even at rather small $t$.

\section{Summary}

In this paper, we have established reliability
of the HAL QCD method by checking systematic uncertainties.
Unlike the direct method, the time-dependent HAL QCD method
is free from the elastic state contamination in two-baryon systems,
and source dependence can be controlled in the derivative expansion.
We have also shown the convergence of the derivative expansion in the non-local kernel,
and the next leading order correction is negligible at low energies.

We have revealed that the fake plateau in the direct method
is caused by the contaminations from low-lying elastic excited states,
and established that finite volume eigenenergies from the HAL QCD potential
agree with effective energies of projected correlation functions.

\section*{Acknowledgements}

The lattice QCD calculations have been performed on Blue Gene/Q
at KEK (Nos. 12/13-19, 13/14-22, 14/15-21, 15/16-12),
HA-PACS at University of Tsukuba (Nos. 13a-23, 14-20) and K computer
at AICS (hp150085, hp160093).
This research was supported by MEXT as ``Priority Issue on Post-K computer''
(Elucidation of Fundamental Laws and Evolution of the Universe) and JICFuS.


\begin{thebibliography}{9}

\bibitem{FVM-review}
  T.~Yamazaki,
  PoS LATTICE {\bf 2014} (2015) 009
  [arXiv:1503.08671 [hep-lat]],
  and the references therein.

\bibitem{Ishii:2006ec}
  N.~Ishii, S.~Aoki and T.~Hatsuda,
  Phys.\ Rev.\ Lett.\  {\bf 99} (2007) 022001 [nucl-th/0611096].


\bibitem{Yamazaki:2011nd} 
  T.~Yamazaki {\it et al.} [PACS-CS Collaboration],
  Phys.\ Rev.\ D {\bf 84}, 054506 (2011)
  [arXiv:1105.1418 [hep-lat]].

\bibitem{Yamazaki:2012hi}
  T.~Yamazaki, K.~i.~Ishikawa, Y.~Kuramashi and A.~Ukawa,
  Phys.\ Rev.\ D {\bf 86} (2012) 074514 [arXiv:1207.4277 [hep-lat]];

\bibitem{Yamazaki:2015asa}
  T.~Yamazaki, K.~i.~Ishikawa, Y.~Kuramashi and A.~Ukawa,
  Phys.\ Rev.\ D {\bf 92} (2015) 1,  014501 [arXiv:1502.04182 [hep-lat]].

\bibitem{Beane:2011iw} 
  S.~R.~Beane {\it et al.} [NPLQCD Collaboration],
  Phys.\ Rev.\ D {\bf 85}, 054511 (2012)
  [arXiv:1109.2889 [hep-lat]].

 \bibitem{Beane:2012vq} 
  S.~R.~Beane {\it et al.} [NPLQCD Collaboration],
  Phys.\ Rev.\ D {\bf 87}, 034506 (2013)
  [arXiv:1206.5219 [hep-lat]].
  
 \bibitem{Beane:2013br} 
  S.~R.~Beane {\it et al.} [NPLQCD Collaboration],
  Phys.\ Rev.\ C {\bf 88}, 024003 (2013)
  [arXiv:1301.5790 [hep-lat]].
  
  \bibitem{Orginos:2015aya} 
  K.~Orginos, A.~Parreno, M.~J.~Savage, S.~R.~Beane, E.~Chang and W.~Detmold,
  Phys.\ Rev.\ D {\bf 92}, 114512 (2015)
  [arXiv:1508.07583 [hep-lat]].


\bibitem{Berkowitz:2015eaa} 
  E.~Berkowitz, T.~Kurth, A.~Nicholson, B.~Joo, E.~Rinaldi, M.~Strother, P.~M.~Vranas and A.~Walker-Loud,
  Phys.\ Lett.\ B {\bf 765}, 285 (2017)
  [arXiv:1508.00886 [hep-lat]].

\bibitem{Wagman:2017tmp} 
  M.~L.~Wagman, F.~Winter, E.~Chang, Z.~Davoudi, W.~Detmold, K.~Orginos, M.~J.~Savage and P.~E.~Shanahan,
  arXiv:1706.06550 [hep-lat].

\bibitem{Aoki:2012tk}
  S.~Aoki {\it et al.} [HAL QCD Collaboration],
  PTEP {\bf 2012} (2012) 01A105 [arXiv:1206.5088 [hep-lat]].

\bibitem{HALQCD:2012aa}
  N.~Ishii {\it et al.} [HAL QCD Collaboration],
  Phys.\ Lett.\ B {\bf 712} (2012) 437  [arXiv:1203.3642 [hep-lat]].


\bibitem{Iritani:2015dhu}
  T.~Iritani [HAL QCD Collaboration],
  PoS LATTICE {\bf 2015} (2016) 089
  [arXiv:1511.05246 [hep-lat]].

\bibitem{Iritani:2016jie}
  T.~Iritani {\it et al.},
  JHEP {\bf 1610} (2016) 101
  [arXiv:1607.06371 [hep-lat]].

\bibitem{Iritani:2017rlk} 
  T.~Iritani {\it et al.},
  Phys.\ Rev.\ D {\bf 96}, no. 3, 034521 (2017)
  doi:10.1103/PhysRevD.96.034521
  [arXiv:1703.07210 [hep-lat]].

\bibitem{Aoki:2017byw} 
  S.~Aoki, T.~Doi and T.~Iritani,
  arXiv:1707.08800 [hep-lat].

\bibitem{Luscher:1991} 
  M.~L\"uscher, Nucl. Phys. B {\bf 354}, 531 (1991).

\bibitem{Aoki:2016} 
  S.~Aoki, PoS LATTICE {\bf 2016} (2016) 109 [arXiv:1610.09763].


\bibitem{Yamazaki:2017euu} 
  T.~Yamazaki {\it et al.} [PACS Collaboration],
  PoS LATTICE {\bf 2016}, 108 (2017)
  [arXiv:1702.00541 [hep-lat]].
 
\bibitem{Savage:2016egr} 
  M.~J.~Savage,
  PoS LATTICE {\bf 2016}, 021 (2016)
  [arXiv:1611.02078 [hep-lat]].
\end{thebibliography}

\end{document}